\documentclass{svproc}
\usepackage{url, graphicx, amsmath}
\usepackage{amsmath,amssymb}
\usepackage{color}
\usepackage{cite}

\begin{document}
\mainmatter              
\title{Finite Element Solution of the Two-Dimensional Bates Model for Option Pricing Under Stochastic Volatility and Jumps}
\titlerunning{HOC--FD vs FEM for Bates PIDE}  
%
\author{Neda Bagheri Renani\inst{1} \and Daniel \v{S}ev\v{c}ovi\v{c}\inst{2}
}
\authorrunning{N.Bagheri Renani,~D.\v{S}ev\v{c}ovi\v{c}} 
%
\tocauthor{}
\institute{Department of Applied Mathematics and Statistics, Comenius University in Bratislava, Slovakia\inst{1,2}; \\
\email{neda.bagheri@fmph.uniba.sk\inst{1}, sevcovic@fmph.uniba.sk\inst{2}}}

\maketitle              

\begin{abstract}
We propose a fourth--order compact finite--difference (HOC--FD) scheme
for the transformed Bates partial integro--differential equation (PIDE).
The method employs an implicit--explicit (IMEX) Crank--Nicolson framework
for local terms and Simpson quadrature for the jump integral. Benchmarks
against second--order finite differences (FD) and quadratic finite elements
(FEM, $p=2$) confirm near--fourth--order spatial accuracy for HOC--FD,
near--second--order for FEM, and second--order temporal convergence for
all time integrators. Efficiency tests show that HOC--FD achieves similar
accuracy at up to two orders of magnitude lower runtime than FEM,
establishing it as a practical baseline for option pricing under stochastic
volatility jump--diffusion models.
\end{abstract}

\keywords{Bates model; option pricing; partial integro–differential equation (PIDE); high order compact finite difference; finite element method}


%
\section{Introduction}

Stochastic volatility models with jumps, such as the Bates model, lead to
partial integro--differential equations (PIDEs) central to option pricing.
Their complexity motivates efficient high--order numerical solvers
\cite{cruz2019illiquid,cruz2020bessel}. Compact high--order finite difference
schemes have been successfully applied to liquidity shock models
\cite{koleva2017fourth}, nonlinear PDEs \cite{koleva2011kernel}, fractional
extensions \cite{dimitrov2016nonuniform}, and uncertain correlation problems
\cite{koleva2014positive}, confirming their advantages for accuracy and stability.  

We develop a fourth--order compact finite--difference (HOC--FD) scheme for the
transformed Bates PIDE, combining an IMEX--Crank--Nicolson split for local terms
with Simpson quadrature for the jump integral. Benchmarks against second--order
finite differences and quadratic FEM demonstrate near fourth--order spatial
accuracy, second--order temporal accuracy, and substantial efficiency gains,
establishing HOC--FD as a competitive baseline for option pricing under
stochastic volatility jump--diffusion models.

\section{Stochastic Differential Equations of the Bates Model}

The Bates model augments the Heston stochastic–volatility framework with a jump component in the asset price, combining continuous variance dynamics with discontinuous returns to better reproduce market skew, fat tails, and smile effects (\cite{bagheri2017comparison,thomee2007galerkin,black1973pricing} see also \cite{cruz2019illiquid}).

\paragraph{Risk–neutral dynamics.}
Let \(S(t)\) denote the price of the asset and \(v(t)\) the variance. Under the risk–neutral measure with correlation \(\rho\) between Brownian drivers,
\begin{equation*}
     dS(t) = \mu S(t) dt + \sqrt{v(t)} S(t) dW_S(t) + S(t) dJ(t), \\
  \end{equation*}  
  \begin{equation*}
    dv(t) = \kappa (\theta - v(t)) dt + \sigma \sqrt{v(t)} dW_v(t),
\end{equation*}
where \(\kappa>0\) (mean–reversion speed), \(\theta>0\) (long–run variance), and \(\sigma>0\) (volatility of variance). The jump process \(J(t)\) is Poisson with intensity \(\lambda\); logarithmic jump sizes \(Y\) are often modeled Gaussian with mean \(\mu_J\) and variance \(\sigma_J^2\), yielding multiplicative price jumps \(S\mapsto S e^{Y}\) \cite{during2019highorder,merton1976}.

\paragraph{Option–pricing PIDE.}
For an option value \(V(S,v,t)\) with a risk-free rate \(r\), the price satisfies the partial integro–differential equation:
\begin{align*}
 & \frac{\partial V}{\partial t} 
    + \frac{1}{2} S^2 v \frac{\partial^2 V}{\partial S^2} 
    + \rho \sigma v S \frac{\partial^2 V}{\partial S \partial v} 
    + \frac{1}{2} \sigma^2 v \frac{\partial^2 V}{\partial v^2} 
    + (r - \lambda) S \frac{\partial V}{\partial S} 
    + \kappa (\theta - v) \frac{\partial V}{\partial v}  \nonumber\\
   & - (r + \lambda) V 
    = \lambda \int_{\mathbb{R}} \left[ V(Se^z, v, t) - V(S, v, t) - 
    \left(e^z-1\right) S \frac{\partial V}{\partial S} (S, v, t) \right] f(z)\, dz.
\end{align*}
where \(f(z)\) is the density of the log--jump size \(z\) (e.g., Gaussian with mean \(\mu_J\) and variance \(\sigma_J^2\)).
where \(f(z)\) denotes the density of the log jump size \(z\) (typically \(z\sim\mathcal{N}(\mu_J,\sigma_J^2)\)) \cite{bates1996,ballestra2010evaluation}.

\subsection{Transformed Bates Model PIDE}

To obtain a numerically convenient form, we introduce the backward time and normalized variables.
\[
\tau = T - t \qquad
x = \ln \frac{S}{K} \qquad
y = \frac{\sigma}{v} \qquad
u(x,y,\tau) = \frac{1}{K}\,e^{(r+\lambda)\tau}\,V(S,v,t).
\]
Here, $K$ is the strike, $r$ the risk–free rate, and the parameter $\lambda>0$ measures the intensity of jumps. Let $f(z)$ denote the
probability density of the logarithmic jump size \cite{pankratov2018nonlinear,during2019highorder}.
With these variables and using differentiation rules, we have:
\[
S\frac{\partial}{\partial S} = \frac{\partial}{\partial x},\ 
v\frac{\partial}{\partial v} = - y \frac{\partial}{\partial y}, \ 
S^2\frac{\partial^2}{\partial S^2} = \frac{\partial^2}{\partial x^2} - \frac{\partial}{\partial x},\ 
v\frac{\partial}{\partial v} = - y \frac{\partial}{\partial y}, 
\]
\[
v\frac{\partial^2}{\partial v^2} =  \frac{y^3}{\sigma} \frac{\partial^2}{\partial y^2}
+ 2 \frac{y^2}{\sigma} \frac{\partial}{\partial y}, \ 
v S\frac{\partial^2}{\partial S\partial v} = - y \frac{\partial^2}{\partial x\partial y}, \ 
\]
\[
\frac{\partial V}{\partial t} -(r+\lambda) V = - K\,e^{-(r+\lambda)\tau}\, \frac{\partial u}{\partial \tau},\ \ 
V(Se^z, v, t) = K \,e^{-(r+\lambda)\tau} u(x+z, y,\tau) .
\]
Thus, the transformed PIDE reads as follows:
\begin{align}\label{eq:transformed-bates}
&  -\frac{\partial u}{\partial \tau} +\frac{\sigma}{2y}\left( \frac{\partial^2 u}{\partial x^2} - \frac{\partial u}{\partial x}\right) -\rho\sigma y \frac{\partial^2 u }{\partial x\partial y}
+\frac{\sigma}{2}\left( y^3 \frac{\partial^2 u}{\partial y^2}+ 2 y^2 \frac{\partial u}{\partial y}\right)
\nonumber\\
&+ (r-\lambda)\frac{\partial u}{\partial x} -\kappa\left(\theta-\frac{\sigma}{y}\right)
\frac{y^{2}}{\sigma} \frac{\partial u}{\partial y} \\
&+\lambda \int_{\mathbb{R}} 
\left[
u(x+z, y, \tau) - u(x, y, \tau) - \left( e^z -1\right)\frac{\partial u}{\partial x} (x,y,\tau)
\right] f(z)\,dz = 0. 
\nonumber
\end{align}



\section{Spatial Discretization and IMEX--Crank--Nicolson}\label{sec:discretization}

We consider a uniform tensor product grid
$\{(x_i,y_j):\ i=0,\dots,N_x,\ j=0,\dots,N_y\}$, 
$x_i = x_{\min}+i\,h_x,\quad y_j = y_{\min}+j\,h_y$,
with spatial discretization steps $h_x,h_y>0$. Let $u_{i,j}^n$ denote the approximation to $u(x_i,y_j,\tau_n)$
at time levels $\tau_n = n k$, where $k = T/N_\tau$. The center difference operators
$D_x, D_y, D_{xx}, D_{yy}, D_{xy}$ will be used for the spatial derivatives (their
concrete stencils are given in Sections~\ref{subsec:compact-local}--\ref{subsec:jump}).
The local (differential) part of (\ref{eq:transformed-bates}) reads as:
\begin{equation*}
\label{eq:local-cont}
L [u]
= \frac{\sigma}{2y}\,u_{xx}
+ \frac{\sigma}{2} y^3\,u_{yy}
- \rho \sigma y\,u_{xy}
+ \left(r-\lambda - \frac{\sigma}{2y}\right) u_x
+ \left(\sigma y^2 + \kappa y - \kappa\theta \frac{y^2}{\sigma}\right) u_y .
\end{equation*}
Its discrete counterpart at $(x_i,y_j)$ is
\begin{align*}
\label{eq:local-disc}
L_h[u]_{i,j}
= \frac{\sigma}{2 y_j}\,D_{xx}u
+ \frac{\sigma}{2} y_j^3\,D_{yy}u
&- \rho \sigma y_j\,D_{xy}u
+ \left(r-\lambda - \frac{\sigma}{2 y_j}\right) D_x u \nonumber\\
+ \left(\sigma y_j^2 + \kappa y_j - \kappa\theta \frac{y_j^2}{\sigma}\right) D_y u .
\end{align*}
The nonlocal (jump) term in (\ref{eq:transformed-bates}) is treated explicitly. Its discrete approximation $I_h[u]_{i,j}$ is designed to match the integral part in (\ref{eq:transformed-bates}):
\begin{equation*}
\label{eq:jump-cont}
I_h[u]_{i,j}=\lambda \int_{\mathbb R}
\big(u(x_i+z,y_j,\tau) - u(x_i,y_j,\tau) - (e^z-1)\,u_x(x_i,y_j,\tau)\big) f(z)\,dz .
\end{equation*}
With these definitions, the semi--discrete system has the form
\begin{equation*}
\frac{ du}{d\tau} u = L_h [u] + I_h [u] .
\end{equation*}
For time stepping, we use an IMEX--Crank--Nicolson scheme: the local operator $L_h$ is
treated implicitly, and the jump operator $I_h$ explicitly. Writing $U^n$ for the vector
of all grid values at time $\tau_n$, one step from $n$ to $n+1$ is given by
\begin{equation*}
\label{eq:imex-cn}
\frac{U^{n+1} - U^{n}}{k}
= \frac12\, L_h  [U^{n+1} + U^{n} ] + I_h[U^{n}] .
\end{equation*}
Sections~\ref{subsec:compact-local} and~\ref{subsec:jump} provide the compact spatial
stencils for $L_h$ and the quadrature/convolution realization of $I_h$, respectively (cf.~\cite{ballestra2010evaluation,during2019highorder,thomee2007galerkin}).

\subsection{Fourth-Order Compact Discretization of the Local Operator}\label{subsec:compact-local}
We denote $u_{i,j} \approx u(x_i,y_j,\tau)$ and adopt fourth–order compact
relations to approximate first-, second-, and mixed derivatives on a $3\times3$
stencil. Such compact schemes are well established for elliptic and parabolic
PDEs with variable coefficients and are known for their spectral-like resolution
\cite{lele1992compact,spotz1996high,during2019highorder,pitkin2020high}.

\paragraph{Compact line relations (1D, fourth order).}
For fixed $j$ (differentiation in $x$),
\begin{align}
\tfrac14 u_{x,i-1,j} + u_{x,i,j} + \tfrac14 u_{x,i+1,j}
&= \frac{u_{i+1,j}-u_{i-1,j}}{2h_x}, \label{eq:compact-ux}\\[4pt]
\tfrac1{12} u_{xx,i-1,j} + \tfrac{10}{12} u_{xx,i,j} + \tfrac1{12} u_{xx,i+1,j}
&= \frac{u_{i-1,j}-2u_{i,j}+u_{i+1,j}}{h_x^2}. \label{eq:compact-uxx}
\end{align}
For fixed $i$ (differentiation in $y$),
\begin{align}
\tfrac14 u_{y,i,j-1} + u_{y,i,j} + \tfrac14 u_{y,i,j+1}
&= \frac{u_{i,j+1}-u_{i,j-1}}{2h_y}, \label{eq:compact-uy}\\[4pt]
\tfrac1{12} u_{yy,i,j-1} + \tfrac{10}{12} u_{yy,i,j} + \tfrac1{12} u_{yy,i,j+1}
&= \frac{u_{i,j-1}-2u_{i,j}+u_{i,j+1}}{h_y^2}. \label{eq:compact-uyy}
\end{align}

\paragraph{Mixed derivative via sequential compact differentiation.}
Let $M_x=\mathrm{tridiag}(\tfrac14,1,\tfrac14)$ and $D_x$ the centered difference
line operator in $x$, with $M_y,D_y$ defined analogously.
The mixed derivative is evaluated by
\[
U_{xy} = M_x^{-1} D_x\big(M_y^{-1} D_y U\big)
        = M_y^{-1} D_y\big(M_x^{-1} D_x U\big).
\]

\paragraph{Compact local operator (nodewise form).}
With coefficients taken directly from the transformed Bates operator
\eqref{eq:transformed-bates}, we have
\begin{equation}\label{eq:compact-local}
(L_h [U])_{i,j} = a_{i,j} U_{xx} + b_{i,j} U_{yy} + c_{i,j} U_{xy}
+ d_{i,j} U_x + e_{i,j} U_y + f_{i,j} U_{i,j},
\end{equation}
where
\begin{align*}
 &  a_{i,j}=\frac12\frac{\sigma}{y_j},\quad
b_{i,j}=\frac12\sigma y_j^3,\quad
c_{i,j}=-\rho\sigma y_j,\quad
d_{i,j}=(r-\lambda)-\frac12\frac{\sigma}{y_j},\quad\nonumber\\
& e_{i,j}=\sigma y_j^2+\kappa y_j-\kappa\theta \frac{y_j^2}{\sigma},\quad
f_{i,j}=0. 
\end{align*}
Here $U_x,U_y,U_{xx},U_{yy},U_{xy}$ are supplied by the compact
relations~\eqref{eq:compact-ux}–\eqref{eq:compact-uyy} and the mixed
construction above. The interior accuracy satisfies
\[
L_h[u] - L[u] = O(h_x^4+h_y^4).
\]

\paragraph{Nine-point compact balance (assembled stencil).}
Eliminating auxiliary derivatives from
\eqref{eq:compact-ux}–\eqref{eq:compact-uyy} within
\eqref{eq:compact-local} yields a $3\times 3$ balance
\[
\sum_{\ell=0}^8 \alpha_\ell u_\ell = \sum_{\ell=0}^8 \gamma_\ell f_\ell,
\]
with $\{u_\ell\}$ the stencil values centered at $(i,j)$ and $\{f_\ell\}$
the right-hand side samples (arising from PDE insertion at $(i,j)$).
The weights $\alpha_\ell,\gamma_\ell$ depend only on the local coefficients
and $(h_x,h_y)$.

\paragraph{Verification template (constant coefficients).}
For $a=b=1$, $c=d=e=0$, $h_x=h_y=h$, the scheme reduces to the
classical fourth-order nine-point Laplacian
\cite{achdou2005partial,cont2005finite}
\begin{align*}
   & \frac{-u_{i-1,j-1}+16u_{i,j-1}-u_{i+1,j-1}
+16u_{i-1,j}-60u_{i,j}+16u_{i+1,j}}{12h^2} \nonumber\\
&+\;\frac{-u_{i-1,j+1}+16u_{i,j+1}-u_{i+1,j+1}}{12h^2}
= f_{i,j} + \tfrac{h^2}{12}\,\Delta f_{i,j}.
\end{align*}
providing a convenient code check in line with established compact
formulations for financial PIDEs \cite{during2019highorder,during2024hedging,
mao2024fem,bagheri2017comparison,balajewicz2017reduced}.

\subsection{Nonlocal Jump Operator and IMEX--CN Time Marching}\label{subsec:jump}

The nonlocal jump part of (1) is
\[
\lambda \int_{\mathbb R}
\big[u(x+z,y,\tau)-u(x,y,\tau)-(e^z-1)u_x(x,y,\tau)\big] f(z)\,dz,
\]
with log-normal jump density $f(z)$
\cite{merton1976,bates1996,cont2005finite}. The compensator
$(e^z-1)u_x$ guarantees integrability and preserves the martingale
condition.

\paragraph{Discrete jump operator.}
On the grid, we approximate
\[
(I_h[u])_{i,j} = \lambda \sum_{m=-M}^{M} \omega_m
\big[u(x_i+z_m,y_j,\tau)-u(x_i,y_j,\tau)-(e^{z_m}-1)D_x u(x_i,y_j,\tau)\big],
\]
using quadrature nodes $\{z_m\}$ and weights $\{\omega_m\}$, with
interpolation for off-grid values
\cite{cont2005finite,during2019highorder,mao2024fem}.

\paragraph{IMEX--CN scheme.}
The semi-discrete system
\[
\frac{d}{d\tau} U = L_h [U] + I_h [U]
\]
is advanced in time by the IMEX--Crank--Nicolson method
\cite{kreiss1970initial,bagheri2017comparison}:
\[
\frac{U^{n+1}-U^n}{k}
= \frac12 L_h[U^{n+1}+U^n] + I_h[U^n],
\]
treating $L_h$ implicitly and $I_h$ explicitly. This combination
retains unconditional stability for the stiff diffusion operator while
avoiding costly implicit treatment of the integral. The scheme fits
naturally into high-order compact PIDE solvers
\cite{during2019highorder,during2024hedging,pitkin2020high}.

\section{Initial and Boundary Conditions}

For the transformed Bates PIDE we prescribe the payoff as the initial
condition at $\tau=0$:
\[
u(x,y,0) = \max(1 - e^x, 0).
\]
At the far left boundary $x_{\min}$, the option value approaches the
discounted intrinsic value. Thus we impose
\[
u(x_{\min}, y, \tau) = e^{(r+\lambda)\tau}\,\big(1 - e^{x_{\min}}\big).
\]
At the far right boundary $x_{\max}$, the put option value decays, so we
set:
\[
u(x_{\max}, y, \tau) = 0.
\]
In the variance direction, we impose homogeneous Neumann conditions:
\[
\partial_y u(x, y_{\min}, \tau) = 0, 
\qquad
\partial_y u(x, y_{\max}, \tau) = 0.
\]
These conditions ensure consistency with the no-arbitrage bounds for
European puts under stochastic volatility jump--diffusion dynamics.

\section{Benchmarking via the Finite Element Method}

We benchmark the compact finite-difference scheme against a conforming finite
element method (FEM) applied to the transformed Bates PIDE under identical
payoff and boundary conditions. The comparison highlights accuracy, stability,
and computational cost.

\paragraph{Weak formulation.}
Writing the local operator in divergence form, we enforce economic Dirichlet
conditions on vertical boundaries and homogeneous Neumann conditions on
horizontal ones, consistent with the FD setup. The weak form reads
\[
\Big(\tfrac{u^{n+1}-u^n}{\Delta\tau},\psi\Big)
+ a\!\left(\tfrac{u^{n+1}+u^n}{2},\psi\right)
= \Big(\tfrac32 L_I[u^n]-\tfrac12 L_I[u^{n-1}],\psi\Big),
\quad \forall \psi\in V,
\]
with $V=\{\psi\in H^1(\Omega):\psi|_{\Gamma_D}=0\}$ and bilinear form
\[
a(u,\psi)=\int_\Omega A\nabla u\cdot\nabla\psi\,dxdy
+ \int_\Omega (b\cdot\nabla u)\psi\,dxdy.
\]

\paragraph{Spatial discretization.}
In a regular shape mesh we use $V_h\subset V$ with continuous quadratic
elements (P2/Q2). The Gaussian quadrature yields the mass $M$, stiffness $K$, and
convection $C$ matrices. For coefficient vector $U^n$,
\[
A_h U^{n+1} = B_h U^n + \Delta\tau\Big(\tfrac32 F^n - \tfrac12 F^{n-1}\Big),
\]
with $A_h=M+\tfrac{\Delta\tau}{2}(K+C)$, $B_h=M-\tfrac{\Delta\tau}{2}(K+C)$,
and $F^n=(L_I[u_h^n],\phi_i)$.\\
For time-independent coefficients, a sparse
factorization of $A_h$ is reused.

\paragraph{Jump treatment.}
The nonlocal operator is evaluated explicitly by Simpson quadrature on a
truncated log-jump interval, with FEM interpolation at off-grid $x$-values.
This keeps the implicit system confined to local terms.

\paragraph{Takeaway.}
FEM achieves near second-order spatial convergence for quadratic elements
\cite{thomee2007galerkin,ballestra2010evaluation,uzunca2020pricing}, but in
structured domains HOC–FD attains comparable accuracy with significantly
lower runtime and memory \cite{during2019highorder,pitkin2020high}.

\section{Numerical Experiments}
We evaluate the (HOC–FD) scheme for the transformed Bates PIDE against second–order finite differences and FEM
with Lagrange bases of degree $p=1,2$. All solvers employ identical payoff
and boundary conditions, the IMEX–CN scheme for local terms,
and explicit Simpson quadrature on a truncated interval for the jump
integral. Accuracy is measured by discrete $L^2$ error and RMSE relative to
a high–resolution reference, while efficiency is assessed in terms of
degrees of freedom, CPU time, memory usage, and a cost–accuracy ratio.

\begin{table}[htbp]
\centering
\footnotesize 
\setlength{\tabcolsep}{4pt}%
\renewcommand{\arraystretch}{0.95}%
\caption{Simulation parameters used in Bates model experiments}
\label{tab:params}
\begin{tabular}{l|r|l|r|l|r}
\hline
\textbf{Parameter} & \textbf{Value} & \textbf{Parameter} & \textbf{Value} & \textbf{Parameter} & \textbf{Value} \\
\hline
$K$ & 110 & $T$ (years) & 1.0 & $r$ & 0.03 \\

$\kappa$ & 1.8 & $\theta$ & 0.02 & $\rho$ & $-0.4$ \\

$\sigma$ & 0.15 & $\lambda$ & 0.25 & Jump law & Lognormal \\

$\mu_J$ & set per exp. & $\sigma_J$ & set per exp. & -- & -- \\
\hline
\end{tabular}
\end{table}

\begin{figure}[h!]
    \centering
`    \includegraphics[width=0.9\textwidth]{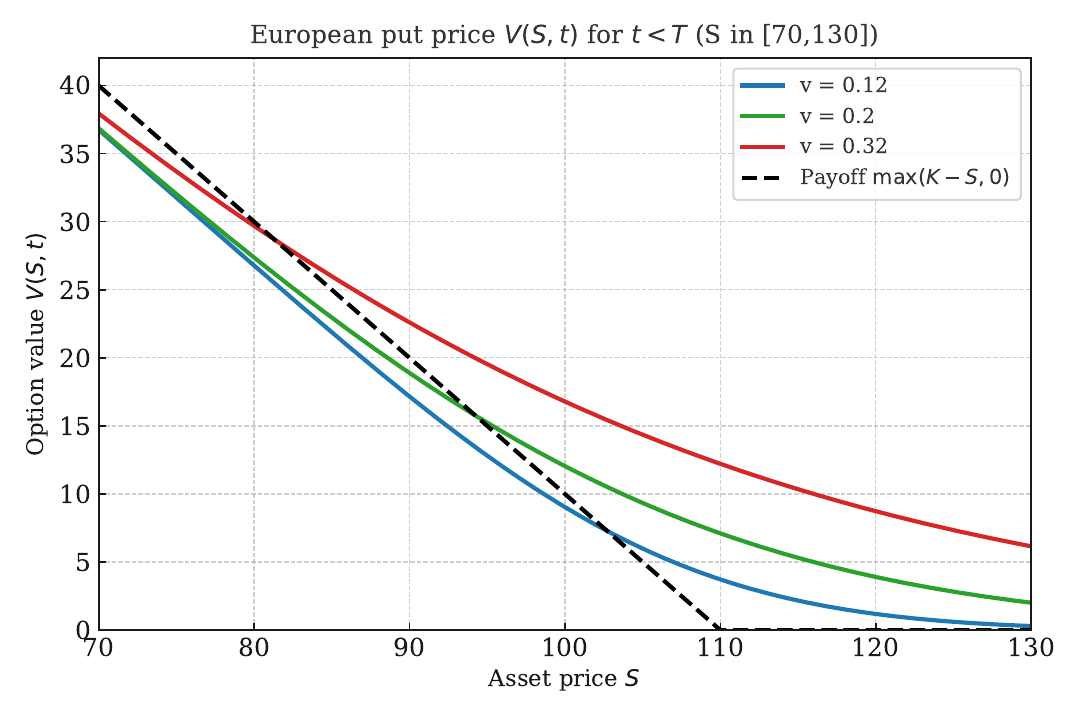} 
   \caption{European put price $V(S,t)$ under the Bates model for 
$S \in [70,130]$. The intrinsic payoff $\max(K-S,0)$ and the 
no-arbitrage lower bound $\max\{Ke^{-rT}-S,0\}$ are shown as 
reference curves}
  \label{fig:bates-euro-put}
\end{figure}

\subsection{Numerical Convergence}

We measure numerical accuracy using three error norms:

\[
\|u^h - u^{ref}\|_{L^2} 
= \left( h_x h_y \sum_{i=1}^{N_x} \sum_{j=1}^{N_y} (u^h_{i,j} - u^{ref}_{i,j})^2 \right)^{1/2},
\]

the discrete $L^2$ error norm, which approximates the continuous integral norm;

\[
\text{RMSE} 
= \left( \frac{1}{N_x N_y} \sum_{i=1}^{N_x} \sum_{j=1}^{N_y} (u^h_{i,j} - u^{ref}_{i,j})^2 \right)^{1/2},
\]

the root-mean-square error, representing average nodal error; and

\[
\|u^h - u^{ref}\|_{L^\infty} 
= \max_{1 \leq i \leq N_x, \, 1 \leq j \leq N_y} |u^h_{i,j} - u^{ref}_{i,j}|,
\]

the maximum pointwise error.

Thus, $L^2$ reflects global accuracy, RMSE captures average nodal error, and $L^\infty$ highlights the worst-case deviation \cite{thomee2007galerkin,bagheri2017comparison,kreiss1970initial}.

\begin{table}[h!]
\centering
\caption{Observed convergence orders in space and time, comparing HOC--FD
with second--order finite differences \cite{cont2005finite,bagheri2017comparison}
and FEM(P2) \cite{thomee2007galerkin,ballestra2010evaluation,uzunca2020pricing}.
Spatial studies fix $\Delta\tau/h^2$; temporal studies fix $h$}

\begin{tabular}{l|c|c}
\hline
\textbf{Method} & \textbf{Spatial order} $p_s$ & \textbf{Temporal order} $p_t$ \\ \hline
FEM (P2)          & $\approx 2.0$ (both $L^2$, $L^\infty$) & $\approx 2.0$ (CN, BDF2, MP) \\ 
HOC--FD (compact) & $\approx 4.0$ (both $L^2$, $L^\infty$) & $\approx 2.0$ (IMEX--CN) \\ \hline
\end{tabular}
\end{table}

\subsection{Computational Efficiency Comparison}

Efficiency was assessed using the cost--accuracy ratio
\[
\eta_{A/B} = \frac{\varepsilon_A^2\,t_A}{\varepsilon_B^2\,t_B},
\qquad \varepsilon \in \{\Vert\cdot\Vert_{L^2}, \mathrm{RMSE}\},
\]
reported relative to the HOC--FD baseline ($\eta=1$). Here $t$ denotes CPU time.

Table~3 shows that, at matched settings, the compact HOC--FD scheme attains
comparable accuracy at substantially lower runtime on structured grids.
Relative to HOC--FD, FEM(P2) is roughly two orders of magnitude more costly,
while second--order FD variants are within single--digit multiples of the
baseline. These results confirm earlier findings that compact FD schemes
achieve significant efficiency gains over FEM in structured domains
\cite{during2019highorder,ballestra2010evaluation,uzunca2020pricing}.

\begin{table}[htbp]
\centering
\footnotesize
\setlength{\tabcolsep}{6pt}\renewcommand{\arraystretch}{1.05}
\caption{Efficiency snapshot at $h=0.1$ relative to HOC--FD baseline
\cite{during2019highorder,pitkin2020high}. Comparisons include second--order
finite differences \cite{cont2005finite,bagheri2017comparison} and FEM(P2)
\cite{thomee2007galerkin,ballestra2010evaluation,uzunca2020pricing}. Errors
measured against a high--resolution reference; $\eta$ uses $L^2$ error and runtime}
\label{tab:eff-full}
\begin{tabular}{l|rrrrr}
\hline
\textbf{Method} & \textbf{DOF} & \textbf{Time (s)} & \boldmath$L^2$\unboldmath & \textbf{RMSE} & \boldmath$\eta$\unboldmath \\
\hline
HOC--FD (IMEX--CN) & 1{,}681 & 1.106 & 0.0230 & 0.0095 & 1.00 \\
FEM ($P_2$)        & 6{,}561 & 23.426 & 0.1522 & 0.0475 & $\approx 1.40\times10^{2}$ \\

CN (2nd--order FD) & 1{,}681 & 2.230  & 0.0341 & 0.0148 & $\approx 2.99$ \\

BDF2 (2nd--order FD) & 1{,}681 & 2.904 & 0.0417 & 0.0181 & $\approx 4.76$ \\

Midpoint (FD)      & 1{,}681 & 3.117  & 0.0425 & 0.0185 & $\approx 5.21$ \\
\hline
\end{tabular}
\end{table}

\newpage
\section{Conclusion}

We developed a fourth–order compact finite–difference scheme for the transformed
Bates PIDE, combining an IMEX–Crank–Nicolson split for local terms with explicit
Simpson quadrature for the jump integral. Benchmarks against second–order finite
differences and quadratic FEM confirmed near–fourth–order spatial accuracy for the
compact scheme and near–second–order for FEM, with all time integrators exhibiting
second–order accuracy. Efficiency tests on structured grids showed that the compact
scheme achieves substantially lower runtimes—up to two orders of magnitude faster
than FEM—while maintaining comparable pricing accuracy. This establishes the
scheme as a competitive baseline for structured–domain option pricing. Future work
will extend the method to American and path–dependent payoffs, adaptive meshes,
broader jump specifications, and accelerated integral solvers.

\section{Acknowledgements}
The first author was supported by Comenius University grant UK/1024/2025. The second author received support from the Slovak VEGA 1-0493-24 agency.

%
%
\bibliographystyle{splncs03}   
\bibliography{paper}


\end{document}